\documentclass[12pt,a4paper,twoside]{scrartcl}
\usepackage{graphicx}
\usepackage{amsmath}
\usepackage{psfrag}
\usepackage[figuresright]{rotating}
\usepackage{fancyhdr}
\usepackage[width=16cm]{geometry} 
\newcommand{\TheAuthor}{}

\def\skiplinehalf{\medskip\\}

\title{Isospin Violation \\ 
in Threshold $\pi $N Scattering}
\author{T. E. O. Ericson
\skiplinehalf
Physics Department, Cern, CH-1211 Geneva 23, Switzerland}
\date{}
\begin{document}
\maketitle
\begin{abstract}
We discuss the electromagnetic corrections to the $\pi N$  scattering lengths generated by  minimal
e.~m. coupling from a knowledge of the low energy expansion of the $\pi N$ elastic scattering amplitude 
as well as from the nucleon and $\Delta $ pole terms, all taken for purely strong interactions. We assume
the heavy baryon limit; the e.~m. and axial  form factors and masses are assumed to have their empirical values, 
such that there is no free parameter.  The different terms have
 a clear physical and intuitive origin. In particular, a large isospin breaking contribution to the 
isoscalar term appears in the elastic  
charged-pion scattering lengths. We attempt a comparison to the results from chiral effective field theory
(EFT) with a physical interpretation of the 
empirical constants in that approach. The results are applied to the energy shift and width of 
the $\pi ^-p$ atom.
\end{abstract}
\section{INTRODUCTION}
In the limit of purely strong interactions the $\pi N$ scattering amplitudes  at threshold 
are fundamental quantities which enter into the discussion of various  problems.
They provide, for example,  a basic test of the Tomozawa-Weinberg chiral relation for the isoscalar 
and isovector scattering lengths in the limit $m_{\pi }=0$
\begin{equation}
a^-=\omega /(8\pi F_{\pi }^2 \label{Weinberg}
)\,\simeq 0.089~m_{\pi }^{-1} ;~~~~~~~a^+\,=\, 0~,
\end{equation}
where $F_{\pi }=93 ~MeV$ is the pion decay constant\cite{TOM66,WEI66} and $\omega =m_{\pi }$ at
 threshold. The empirical
isovector scattering
length is the main ingredient and uncertainty in the forward dispersion relation,
by which the $\pi NN $ coupling constant $14.11\pm 0.05$ is determined\cite{ERI02}  etc..
The major precision source for these quantities are the remarkable measurements of the
 1s level shifts in pionic hydrogen and deuterium as well as the corresponding widths.
 In the case of pionic hydrogen, present experiments have, or will shortly achieve, a
precision of $0.2\%$ for the shift and $1\%$ for the width \cite{GOT03,PSI2} and these quantities convert in 
principle to similar precision for the scattering amplitudes.

There are 2 major ways by which one can approach the problem of determining the $\pi ^-p$  scattering
length from pionic hydrogen data.  The first is  chiral effective 
field theory (EFT) \cite{FET01,FET01a,JG02}.  Here one starts from an effective Lagrangian in the
 chiral limit
 and makes first 
a systematic expansion in orders ${\cal O }(p^n)$ of momentum corrections.  The e.~m. contribution to order 
 $\alpha $ and the strong
symmetry breaking  are then expanded as  small additional corrections. This approach introduces
empirical constants to summarize short range contributions and these must be determined from experiments.

The other major approach, which I will follow below,  is complementary and less
 ambitious~\cite{ERI04,ERI05}. It starts from the empirical low energy expansion of the strong interaction
 amplitude in terms of  energy and momentum. It  is rather well explored including its
principal dynamic features, although the experimental  value of the scattering length needs further tuning. 
This system has well defined e.~m. and axial form factors of the
pion and nucleon ($\Delta $) which are experimentally  known as are  the physical masses.
 Starting from this knowledge, we  determine the e.~m. corrections.  
The main guiding tool is minimal e.~m. coupling or, in other words, current   conservation.
In addition, we use the same minimal coupling principle to determine the dispersive contributions from 
the  radiative capture processes dominated by 
the nucleon and $\Delta $ isobar intermediate states.
For this last problem, it is a considerable simplification to work in the heavy baryon
 limit and we will do so below.
 This approximation is also used in the EFT approach. 

The objective of this second approach is to obtain an intuitive  picture of the 
corrections and the physical mechanisms of isospin breaking. It will become clear that
 kinematic considerations in the wide sense is the key to several of the effects.

The plan is the following. I  first discuss the nature and physics of the corrections to the 
leading order Deser-Trueman formula \cite{DES54,TRU61}, which relates the  energy shift to the scattering length.
I  then discuss 'inner' e.~m. corrections to the s wave scattering of
 charged pions from the nucleon at threshold and show that the dominant term is intimately 
connected to the well established p-wave $\pi N$ scattering physics and the $\Delta $ isobar.
This  contribution can summarized by the EFT chiral parameter $f_1$~\cite{FET01,FET01a,FET99}. 
Finally, I make a tentative comparison with the results of the EFT approach.

\subsection{Step 1: How to get scattering lengths from atomic energy shifts: 
removal of the external Coulomb field}
The s-wave threshold amplitude for the strong interaction  has in the case of a single channel
the low energy expansion
\begin{equation}
tg \delta = a_h\,+\,b_hq^2\,+{\cal O}(q^2) \label{Eq:1}
\end{equation}
where $q$ is the momentum and $a_h,b_h$ are the scattering length and the range term respectively.
Such an expansion does not by itself assume isospin invariance, but it means implicitly
 a short ranged interaction.

The atom is highly non-relativistic. To leading order the strong interaction shift
 is obtained using  the $1s$ wave function for the point Coulomb field
\begin{equation}
\phi _{Bohr}(r)=\phi _{Bohr} (0)\exp (-\alpha mr)\simeq \phi _{Bohr}(0)( 1-\alpha mr\,+..),
\end{equation}
where $m$ is the reduced mass and $\alpha mr<<1$ over the strong interaction region.
If we assume that the amplitude of Eq. 
(\ref{Eq:1})  results from a short ranged pseudo-potential ('effective interaction Lagrangian'),
the leading order $1s$ strong interaction energy shift is simply the Born approximation term:
\begin{equation}\epsilon _{1s}=-{4\pi \over 2m}\phi (0)^2\,a_h\,. \label{Eq:2}
\end{equation}
This is the time-honored standard reference shift referred to as the
 Deser-Trueman formula~\cite{DES54,TRU61} and 
there are of course corresponding expressions for any $ns$ state.
We now generalize this expression.

-Assume that the Coulomb field without strong interactions to be generated by the
 extended charge distributions of the pion and the proton.  The corresponding e.~m.
 form factor $F_{em}(q^2)=F_{\pi }(q^2) F_{p}(q^2)$ is  experimentally known.
 Its origin is in principle irrelevant,
although the issue of a Dirac versus a Pauli form factor of the proton 
must be faced later, when we compare to
results of EFT in the heavy baryon limit. At this stage the Pauli form factor is the relevant one..

-This problem is highly non-relativistic and well described using  wave functions.
 Outside the extended charge, the exact solution  is  the Whittaker function. 

-In the limit that the strong interaction (\ref{Eq:1}) has a negligible range  compared to that of
the charge distribution, the corresponding strong interaction energy shift has an 
exact solution~\cite{ERI04}.

-For the energy dependence of the scattering length we use the 'minimal e.~m.
coupling principle" or gauge invariance and replace the energy of a charged particle 
$\omega \to \omega -eV_C(r)$.  Another way of saying the same thing is that one should
locally at the interaction point use the correct interaction energy in the Coulomb potential.

There are now 3 separate physical effects \cite{ERI04}.
First, the starting wave function at the origin is improved using  the Coulomb potential
 corresponding to the
joint charged distribution of the pion and proton, which is simpler to handle than the
 singular point Coulomb interaction.
 Second, the interaction does not correspond to the free scattering 
at threshold of Eq.(\ref{Eq:1}), but corresponds to the energy shifted to that of the
Coulomb potential at the interaction
Third,  to second order in the scattering length,
 the binding gives
characteristic cusp term which is nearly independent of assumptions. 
Finally, we always include tacitly a small, model independent vacuum polarization correction, 
which is conceptually irrelevant in the present context. 

The extended charge changes the Coulomb
 wave function,  chopping off the singular linear behavior of the wave
function near the origin, which comes from the $1/r$ behavior of the Coulomb interaction.
The wave function then varies normally with $r^2$ at
 the origin. The wave function at the origin is then 
\begin{equation}\label{wfchange}
\phi _{Bohr}(0)\to \phi _{Bohr}(0)[1-\alpha m\langle r\rangle _{em}+ {\cal O}(\alpha ^2)]
\end{equation}
 Here  $\langle r\rangle _{em}$ is the expectation value over the extended charge density. 
The corresponding change in the energy  shift is
 \begin{equation}\label{wfshift}
\delta \epsilon _{1s;wf}=-{4\pi \over 2m}\phi _{Bohr}(0)^2(-2\alpha m \langle r\rangle _{em})
\,a_h\,\simeq (-0.9\%)\,\epsilon _h
\end{equation}
Detailed numerical values are given in Table~1.
\begin{table*}[htb]
\caption{Coulomb corrections in percent.
The vacuum polarization contribution is included in the total correction. 
(From Ref.\cite{ERI04})}
\label{table:1}
\newcommand{\m}{\hphantom{$-$}}
\newcommand{\cc}[1]{\multicolumn{1}{c}{#1}}
\renewcommand{\arraystretch}{1.2}
\begin{tabular}{@{}lccrr}
\hline
&Extended charge           & \cc{Renormalization} & \cc{Gauge term}  & \cc{Total} \\
\hline
$\delta_{1s}$      & $-$0.853(8)& 0.701(4) & $-$0.95(29) &  $-$0.62(29) \\
$\delta_{\Gamma} $    & $-$0.427(4) &  0.701(4) & 0.50(23) &   1.02(23) \\
$\delta_{\pi ^+p\rightarrow \pi ^+{\rm p}}$   &  $\ \ $0.853(8)&   0.72(5)&  $-$1.71(29)&   0.35(29)\\
\hline
\end{tabular}
\end{table*}
The wave function modification  is not specific to the atomic bound state problem.
It has a near identical counterpart in the elastic $\pi ^+p$ scattering close to threshold,
 but the interaction is now repulsive.  For a neutron target such Coulomb corrections are of course absent. 
  we should change the sign
 of the correction term Eq. (\ref{wfchange}), but  there is no
 change for a neutron target, of course. 

The energy change in the scattering amplitude (\ref{Eq:1}) depends only on minimal coupling, such that it happens for the scattering near threshold of a positive pion as well, but with opposite sign.
\begin{equation}
\omega \to \omega -e(1+\tau _3)t_3V_C(r)
. \end{equation}
For a short ranged limit  we must use the  potential energy at the origin.

The range term $b_h$ in Ref. (\ref{Eq:1}) corresponds to the experimental
 isoscalar and isovector range terms $b^+\pm b^-$ for $\pi ^{\mp}p$ interactions with the corresponding $\pi ^-p$ energy shift:
\begin{equation}\label{gaugeshift}
 \delta \epsilon _{1s;gauge}=-{4\pi \over 2m}\phi _{Bohr}(0)^2(-\alpha \langle {1\over r}\rangle _{em})
b^{\pi ^-p}_h\simeq (-1.0\%)\,\epsilon _h~.
\end{equation}
We note first, that it does not matter, to leading order,
 whether this range term is derived in Eq. (\ref{Eq:1}) from terms of the momentum $q$
or from the energy 
$\omega $: the result will be identical~\cite{ERI82}, for the interaction is nearly on
 the mass shell.
Second, the range term $b_h$ has its own physics and should not be taken to be 
 proportional to the scattering length.

The third contribution is the cusp effect generated by rescattering to second order
in the scattering length
by the binding in the Coulomb potential. It depends only weakly on the form factors.
 Its physics is that the incident wave in the interaction is renormalized by the scattering 
(effective field effect) as is known in the present context since a long time. It is accurately
\begin{equation}
\delta \epsilon _{1s;renorm}=-8\pi \alpha a_h^2 ~[2-\gamma +\log 2\alpha -\langle \log mr \rangle _{em}]
\phi _{Bohr}(0)^2\simeq (+0.7\%)\,\epsilon _h
\end{equation}
These corrections are exact to order ${\cal O}(\alpha ^2)$ 
in the limit of a short ranged strong interaction.

A crucial term is the second one corresponding to the energy shifted amplitude depending on the $\pi ^-p$
range parameter $b_h$. One might think that the main contribution would follow 
from the $\omega $ dependence of the dominant isovector Tomozawa-Weinberg scattering
length (\ref{Weinberg}), which indeed appears to generate corresponding terms to next-to-leading  
order in the EFT expansion
\cite{FET01,JG02,FET99}.  This is not the case. This isovector range term is largely canceled 
by the nucleon pole term and the net effect is quite small. 
 Instead the contribution is mainly  generated
by the isoscalar range term proportional to $\omega ^2$.  This term is proportional to $\omega $.

It is a fair question to ask how
accurate the expression for this energy correction is   in practice. This can be inferred in several ways. 
 An easy and intuitive estimate is obtained modifying
the level displacement expression of  Eq.~(\ref{Eq:2})  taking the 
 interaction density to have a form factor
 $F_{str}(r)$ instead of a point interaction.  This means that the Coulomb interaction 
should be averaged over the
interaction region folding in this additional form factor. The corresponding interaction shift
 becomes
$\delta \epsilon _{1s;gauge}\propto \int F_{str}(q^2)F_{\pi }(q^2)G_p(q^2)/q^2\,d^3q$.
 Using standard monopole and dipole  form factors for the pion and proton
with the scale of the $\rho $-meson and observing that the form factor of the
 Tomozawa-Weinberg term is also a monopole associated with the $\rho $-meson, indicates 
a typical small uncertainty from this source of about $0.15\%$, which is beyond present
experimental uncertainties.

Up to this point we have not  used the heavy baryon approximation.  
This means that for the proton charge form factor we should here use the Pauli form factor,
which includes the $1/M_p$ terms from the magnetic charge density and not the Dirac form factor.
 More about that later.

All of these considerations are of course quite general and apply with small modifications 
for any hadronic atom.
They are easily generalized to states $ns$ of higher quantum number $n$ as well as to the width. 
It is also easy to generalize the result to a repulsive Coulomb interaction such as $\pi ^+p$, but
 the results will then concern the scattering length at threshold.  Finally they can be 
generalized to the situation of two coupled channels~\cite{ERI04}. The problem of
extracting the hadronic scattering length from the atomic $\pi ^-p$  energy 
shift is therefore solved to 
a precision of about $0.1\%$ on the level of the Coulombic interaction.
\subsection{Step 2. Transverse photons; a dominant isospin breaking mechanism at threshold}
The previous Coulombic contributions correspond to longitudinal photons.  The question is then of the
importance of transverse photons.
In fact, these generate a large isoscalar correction, i.~e., a term which
 is the same for all the elastic charged pion amplitudes as
 Dr. A.~N.~ Ivanov and myself have recently shown~\cite{ERI05}.

A guide to the importance of such terms is the dominant contribution of the Kroll-Ruderman radiative capture
process $\pi ^-p\to \gamma n$ at threshold~\cite{KRO54} (see Fig.~1a), which experimentally gives
 a $1s $ width of 8\% of the strong
interaction shift~\cite{SPU77}, a huge number. The term we consider is the corresponding dispersive shift
 (see Fig.~1b).

\begin{figure}
\psfrag{p}{$p$}
\psfrag{m}{$\pi ^-$}
\psfrag{a}{$(a)$}
\psfrag{b}{$(b)$}
\psfrag{pm}{$\pi = \pi^{\pm}$}
\psfrag{N}{$ N = p, n$}
\psfrag{g}{$\gamma$}
\psfrag{n}{$n$}
\psfrag{X}{$ N,\Delta$}
 \centering
\includegraphics[height=0.10\textheight]{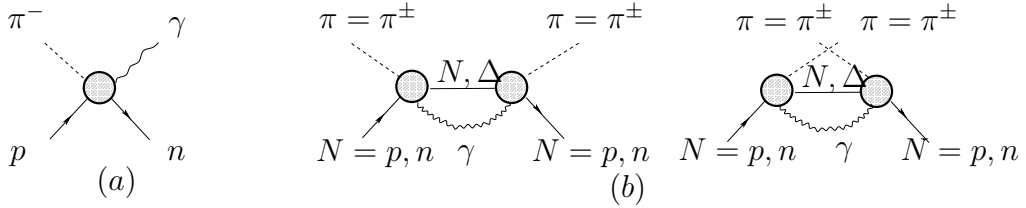}
\caption{
 (a) the Kroll-Ruderman radiative capture process (b) diagrams of  $\pi N \to X \gamma \to \pi N $
  reactions  generating isospin breaking in low--energy elastic $\pi N$ scattering.}
\end{figure}
The matrix element for radiative capture can, for example, be derived from the Partially Conserved Axial Current (PCAC)
 relation using minimal e.~m. coupling (or directly from the nucleon pole term).
\begin{eqnarray}\label{PCAC}
\partial _{\mu }A_{\mu }=-m_{\pi }^2F_{\pi }^2\phi _{\pi }(x);
~\partial _{\mu } \to \partial _{\mu }\pm ie{\cal A}_{\mu },
\end{eqnarray}
where ${\cal A}_{\mu }$ is the e.~m. 4-vector potential.
This statement corresponds to saying that the we have electric dipole (E1) radiation 
due to
 the discontinuity 
in the current or, in other words, a kind of transition radiation.
We calculate the dispersive contribution in the heavy baryon approximation.
 This is particularly
convenient, since in this limit and at threshold  the  radiation comes only from the vertex 
itself  in the Coulomb gauge and not
 from the pion and nucleon.

The characteristic features are: 

-the transition is an axial one, such that its strength is well defined.

-it is natural to use the axial  form factor
$F_A(\vec{q}^{\;2})$, which is empirically well approximated by a dipole shape (see, e. g., \cite{TE88,NAG79})
\begin{eqnarray}\label{label13}
 F_A(\vec{q}^{\;2}) =  (1 + \vec{q}^{\;2}/M^2_A)^{-2}~ {\rm with }~ M_A = (960\pm 30)\,{\rm MeV}
\end{eqnarray}

- typical energy denominators $(p\pm m_{\pi })^{-1}$ appear from intermediate
 states with the sign switching due to crossing.

-More precisely, the  $\pi ^-p$ amplitude contribution  gives a large nucleon isoscalar term of
 3\% in the limit $m_{\pi }=0$.
The contribution is for the $\pi ^-p$ case 
\begin{eqnarray}\label{label15}
  \hspace{-0.3in}&&\Big(1+ \frac{m_{\pi}}{M_N}\Big) \delta f_{1s}^{(n\gamma)} = 
  \frac{3\alpha}{8\pi^2}
  \frac{g^2_A}{F^2_{\pi}}
  \, {\cal P}\int^{\infty}_0 \frac{dp\,p\,F^2_A(p^2) }{p -
    m_{\pi}-i0}.
\end{eqnarray}
However, from the SU3 symmetry point of view, the $\Delta $ isobar and
the nucleon are basically identical, but for the $N\Delta $ mass splitting and weight factors. It would be unnatural to include the nucleon only.
In the limit of no $N\Delta $ mass splitting, the inclusion of  $\Delta $ intermediate states gives a 
multiplicative factor $25/9$ as compared to the
nucleon term. The increases the previous 3\% to 9\%, an enormous correction!

Why is this so large?  The reason is that the scale parameter is the axial mass $M_A$ 
and not the pion one $m_{\pi }$ which gives a factor 7 enhancement with respect to na\"ive expectations.
When the $N\Delta $ mass splitting of $\approx 2m_{\pi }$ is brought in, it cuts the $\gamma \Delta $ term
by $50\%$ to a total isoscalar correction of about $+6\%$. This is still very large, but we expect
relativistic kinematic factors to cut it additionally to about 4.8\%.

\subsubsection{The case of $m_{\pi }\neq 0$}

In this situation new characteristic terms appear of the type $c\,m_{\pi }\ln (m_{\pi }/M_A)+d\, m_{\pi }$,
 which are
generated both from the nucleon and the $\Delta $ intermediate states.  These terms depend only weakly on 
the exact value of $M_A$. In the particular case of the nucleon intermediate state, the term proportional to
$m_{\pi }\ln m_{\pi } $ has the identical coefficient to the one found to third order in 
chiral EFT by Gasser et~al.~\cite{JG02}.
If we  expand  the amplitude $\delta f^{(n\gamma)}$ in
 Eq.~(\ref{label15}) in terms of the small parameter $x~=~x_{\pi}~=
m_{\pi }/M_A$~, we have in this case:
 \begin{eqnarray}\label{label19}
   \hspace{-0.3in}&&\Big(1+ \frac{m_{\pi}}{M_N}\Big) \delta f^{(n\gamma)} =
  \frac{3\alpha }{8\pi ^2}\frac{g^2_A}{F^2_{\pi }}
  \Big[ \frac{5\pi }{32}M_A
   -  m_{\pi }\Big(\ln \frac{m_{\pi }}{M_A} + \frac{11}{12} +
{\cal O}({m_{\pi } \over M_A}\Big)
\Big].
\end{eqnarray}

When the $\Delta $ isobar is degenerate with the 
nucleon,  this nucleon term  strongly canceled by nearly a magnitude, such that the
the dependence on the pion mass becomes  negligible. However,
 when the $N\Delta $ mass splitting
 is introduced in accordance with observation, the $\Delta $ contributions 
 are quenched such that the contribution from the nucleon term is partially restored. 
 A small term in the pion mass of about 50\% of the value for the nucleon only survives.
Numerical values for the different cases are given  in Table 2.
  \begin{table*}  
 \caption{Contributions to the $\pi N$ scattering lengths from  dispersive 
 radiative capture with nucleon and
 $\Delta $ intermediate states in units of $10^{-3}\,m_{\pi }^{-1}$ (from Ref. \cite{ERI05}). }
\label{table:2}
\newcommand{\m}{\hphantom{$-$}}
\renewcommand{\arraystretch}{1.5}
\begin{tabular}{l|lll}
  \hline
$m_{\pi} = 
  0, \omega_{\Delta} = 0$&$(3.0(1)_{N\gamma}
  + 5.3(2)_{\Delta\gamma})\,t^2_3 $&$= 8.3(3)\,t^2_3$ \\
$m_{\pi}
  = 0, \omega_{\Delta} \neq 0$&$(3.0(1)_{N\gamma} + 2.4(1)_{\Delta\gamma})\,t^2_3$ &$=  5.4(2)\,t^2_3$\\
$m_{\pi}
  \neq 0, \omega_{\Delta} = 0$&$(2.6(1)_{N\gamma} + 4.6(1)_{\Delta\gamma})\,t^2_3$ $+ (- 0.8_{N\gamma} + 
 0.7_{\Delta\gamma})\,t_3\tau_3$&$= 7.2(2)\,t^2_3 - 0.1\,t_3\tau_3$\\
$m_{\pi} 
  \neq 0, \omega_{\Delta} \neq 0$&$(2.6(1)_{N\gamma} + 2.5(1)_{\Delta\gamma})\,t^2_3$ $+(-0.8_{N\gamma} +
  0.3_{\Delta\gamma})\,t_3\tau_3$&$= 5.1(2)\,t^2_3 -0.5\,t_3\tau_3$\\
  \hline
\end{tabular}
\end{table*}

\subsection{Step 3. E. M. isospin violation and Chiral EFT}
In the last few years a scheme has been developed to calculate strong and e.~m. isospin violation to leading order
for $\pi N$ scattering using field theory methods based on an effective chiral Lagrangian~\cite{FET01,FET01a,JG02,FET99}
The amplitudes are calculated in a systematic expansion in powers of momenta (EFT). I will not enter into the
 details of this expansion, but only  sketch a tentative comparison of  some specific points. 
It is important to realize that certain of the
predictions of such an effective field theory are specific and outside our present approach, while,
 on the contrary, 
our approach here generates terms of higher order in the EFT description than those presently considered.
In the heavy baryon limit, the e.~m. isospin breaking in the $\pi N$ threshold amplitudes are related to the
the e.~m. mass of the nucleon and the $np$ e.~m. mass difference in the EFT beyond the purely kinematic
 effects~\cite{FET01}.
 To next-to-leading order  these relations can be expressed in terms of 3 constants $f_{1,2,3}$.
 In the case of $\pi ^{\pm }p$ elastic scattering, which I chose for illustration, this gives the following
relations. 
\begin{eqnarray} \label{f12}
M_n^{em} =-e^2F_{\pi }^2\Big[f_1+f_3\Big]~ ;~M_p^{em} =-e^2F_{\pi }^2\Big[f_1+f_2+f_3\Big]~;~~~~
a_{\pi ^{\pm }p}^{em}= -2\pi \alpha [f_1\pm {1\over 4}f_2]
\end{eqnarray}
Following Ref. \cite{FET01}, I omit the term generated by the physical mass difference between the charged
 and neutral pion,
 which is of no concern in the present context. 

 It is of considerable interest to 
attempt to identify our results in the EFT expansion.
Can we match our previous results to its explicit terms? A problem occurs  in view
 of the basic difference with our  approach. For example, in  the discussion of the Coulombic
terms, we start from the
scattering amplitude and form factors as they would result in a description  to all orders, but in
the absence of e.~m. interactions; in this sense we include physics not presently included 
  in the EFT approach.
We must therefore make sure we compare comparable quantities.
First, we must use form factors in the heavy baryon limit. This means that we must use  
the Dirac form factor
$F_p(q^2)$ and not the  Pauli form factor $G_p(q^2)$; the charge distribution generated by the magnetic
moment of order $M_N^{-1}$ should be omitted. We should also to this order omit the wave function correction 
$\alpha m\langle r\rangle _{em}\,a_{\pi ^-p}$ of Eq. (\ref{wfshift}), for the mass scale is set by the $\rho $-meson and it is of order
$\alpha m/m_{\rho}a_h$ or formally of 4th order in the EFT expansion\footnote[1]
{Since its coefficient is large,
its order is unclear; its magnitude corresponds to 3rd order.}.
The relevant term for the comparison is generated by
 Eq.~(\ref{gaugeshift}).  Here the dominant contribution comes from the isoscalar range term
$b^+$. It generates a contribution of order $m\alpha  \langle r\rangle _{em}\,b^+$, which is of 4th order in the
EFT expansion and outside the present EFT discussion; it will require an additional  EFT constant. Similarly, the 
pole term in the isovector range term vanishes in the heavy baryon limit.

\subsubsection{The EFT constant $f_2$, the $np$ mass difference and the Coulomb interaction}
In 
the heavy nucleon limit, the constant $f_2 $ in Eq. (\ref{f12})  describes the $np$ e.~m.
 mass splitting, which  then results from
the Coulomb self energy of the proton with the Dirac form factor:~
$(M_p-M_n)^{em}\propto f_2\propto {1\over 2}\int d^3q [F_p(q^2)]^2/q^2$.
 The corresponding effect in the $\pi ^cp$ scattering must be taken in the same limit with the  Dirac  factor.
 To leading order, the strong scattering amplitude is given
by the Tomozawa-Weinberg term (\ref{Weinberg}), which is linear in  $\omega $.  
The minimal coupling procedure
 generates an isospin-breaking contribution:
\begin{equation}
a_{\pi ^{\pm } p}^{em}= {1\over 2}(1+\tau _3)t_3^2\, {eV_C(0) \over (8\pi F_{\pi }^2)}~,
\end{equation}
where $eV_C(0)\propto \alpha \int F_{\pi }(q^2)F_p^2(q^2)/q^2d^3q$. Compared to the 
proton e.~m. self energy   $f_2$ above, the  similarity is striking and numerically the expressions are equal to about 15\%. However, if the shape of
 the strong interaction is included with an interaction form factor, one obtains equality if
\begin{equation}
F_{str}(q^2)F_{\pi }(q^2)F_p(q^2)\simeq F_p^2(q^2)
\end{equation}
 or, since both the nucleon and pion form factors are governed by the $\rho $-meson mass
$$F_{str}(q^2)\simeq F_{\pi }(q^2)\simeq (1+q^2/m_{\rho }^2)^{-1}.$$
We here obtain unexpectedly a quantitative result for the isovector interaction form factor.
The Tomozawa-Weinberg term is frequently associated with a monopole interaction with a range given by the
 $\rho $ meson mass $(1+q^2/m_{\rho }^2)^{-1}$\cite{BRO75}, while both the nucleon and the pion have form factors 
closely approximated by a dipole, respectively a monopole, form factor with the $\rho $-meson mass.

This is suggestive, since it indicates that, indeed, the EFT $f_2$ coefficient comes from the
Coulomb interaction. On the other hand,
 the EFT  does not presently give the shape of the form factor. The last relation implies
however that there is  an intrinsic link between the proton, pion and the Tomozawa-Weinberg interaction 
form factors in EFT.

A proper account for 
the Coulomb interaction  requires realistically that one  goes beyond the present level of the EFT
 expansion and  accounts for the  Pauli form factor $G_{E;p}$ of the proton as well
as for the terms generated by the isoscalar range term.
\subsubsection{The EFT constant $f_1$ and the axial form factor}
The dispersive contribution from intermediate 
$\gamma N (\Delta )$ states to the scattering length as given in Eq. (\ref{label15}) 
 is isoscalar in the charged pion sector in the
limit of a vanishing pion mass.  This is exactly the symmetry property of the
contribution 
in EFT by the next-to-leading order constant $f_1$. This constant  also appears as a part 
of the e.~m. neutron mass
$M_n^{em}$, but then  it always comes in  the combination $f_1+f_3$~ (see Eq.\ref{f12})).
These terms cannot be physically separated~\cite{FET01}. Here we have an interesting situation.
The neutron e.~m. mass is expected to be quite small compared to the $np$ e.~m. mass splitting 
of about $0.8~MeV$ and  to be dominated by the magnetic self energy. It then vanishes
in the heavy baryon limit with a likely uncertainty of $\pm (0.1~\tilde ~0.2)~MeV$ ~\cite{GAS75}.
This corresponds to a value $F_{\pi }^2|f_1|\simeq 1~MeV$, while our value with the
 physical $\Delta $
isobar included gives $F_{\pi }^2f_1=- 26(1)~MeV$, which is over a magnitude larger.
 Dimensional estimates 
inside of EFT give intermediate estimates $F_{\pi }^2|f_1|= 6~MeV~{\rm and }~12 ~MeV$~\cite{FET01,JG02}.
There appears therefore numerically to be little relation of this parameter with the neutron e.~m.
mass, which suggests a massive cancellation between the  EFT constants $f_{1  }$
 and $f_{3  }$.    Such a cancellation occurs explicitly in a model evaluation inside a heavy quark
 model \cite{LYU02}, which has certain similarities with our description.
The connection of the constant $f_1$ to the nucleon e.~m. mass is therefore tenuous and 
of little practical importance.
\section{Conclusion}
Isospin violation in $\pi N$ elastic scattering has been shown here to have  well determined contributions
originating in the Coulomb field of the extended charge with little model dependence. These corrections 
are general and involve terms beyond present EFT approaches. To leading order in the strong interactions
corresponding terms have counterparts in  EFT in the heavy baryon approximation. 
In addition, terms with $\gamma N  (\Delta )$
 intermediate states
give rise to important isoscalar isospin breaking terms which can be looked at as model descriptions of
the unknown EFT constant $f_1$.  For a non-vanishing pion mass, the same mechanism  generates small, 
model insensitive, isospin breaking in the isovector interaction dependent on the pion mass.

An important finding is that the isospin breaking is small in the isovector amplitude.
This is consistent    with the finding of Meissner et al. \cite{FET01}, but it is
 in violent disagreement with
the important violation reported by Matsinos \cite{MAT97}, which depends only on isospin breaking in the isovector amplitude,
although the author does not explicitly state so.  The Matsinos results have also been shown 
to be  grossly at variance with
the empirical scattering lengths deduced from pionic hydrogen and deuterium~\cite{ERI02}.

Although therefore the chiral EFT constants $f_{1,2}$ can be rather well understood from
physical effects, this is not the whole 
story. Our terms include the main part, but not all, of the isospin breaking mechanisms.
In addition to the effects discussed here, the effective field theory generates generic isospin breaking beyond our
considerations. These  additional terms appear in charge exchange and $\pi ^0N$ elastic scattering.
 They are are conceptually important non-trivial chiral predictions. Numerically, 
such contributions are of the order of 1\% and of similar magnitude as the terms we consider.
 Obviously a quantitative test of the predictions
of EFT requires that that these intrinsic terms  be reliably separated from those which are generated by 
the general mechanisms  discussed here. A meaningful analysis will require both approaches.

\end{document}